\documentclass[twoside,twocolumn,9pt]{article}
\usepackage{extsizes}
\usepackage[super,sort&compress,comma]{natbib} 
\usepackage[version=3]{mhchem}
\usepackage[left=1.5cm, right=1.5cm, top=1.785cm, bottom=2.0cm]{geometry}
\usepackage{balance}
\usepackage{mathptmx}
\usepackage{sectsty}
\usepackage{graphicx} 
\usepackage{lastpage}
\usepackage[format=plain,justification=justified,singlelinecheck=false,font={stretch=1.125,small,sf},labelfont=bf,labelsep=space]{caption}
\usepackage{float}
\usepackage{fancyhdr}
\usepackage{fnpos}
\usepackage[english]{babel}
\addto{\captionsenglish}{%
  
}
\usepackage{array}
\usepackage{droidsans}
\usepackage{charter}
\usepackage[T1]{fontenc}
\usepackage[usenames,dvipsnames]{xcolor}
\usepackage{setspace}
\usepackage[compact]{titlesec}
\usepackage{hyperref}

\usepackage{epstopdf}

\definecolor{cream}{RGB}{222,217,201}

\begin{document}

\pagestyle{fancy}
\thispagestyle{plain}
\fancypagestyle{plain}{
\renewcommand{\headrulewidth}{0pt}
}

\makeFNbottom
\makeatletter
\renewcommand\LARGE{\@setfontsize\LARGE{15pt}{17}}
\renewcommand\Large{\@setfontsize\Large{12pt}{14}}
\renewcommand\large{\@setfontsize\large{10pt}{12}}
\renewcommand\footnotesize{\@setfontsize\footnotesize{7pt}{10}}
\makeatother

\renewcommand{\thefootnote}{\fnsymbol{footnote}}
\renewcommand\footnoterule{\vspace*{1pt}%
\color{cream}\hrule width 3.5in height 0.4pt \color{black}\vspace*{5pt}} 
\setcounter{secnumdepth}{5}

\makeatletter 
\renewcommand\@biblabel[1]{#1}            
\renewcommand\@makefntext[1]%
{\noindent\makebox[0pt][r]{\@thefnmark\,}#1}
\makeatother 
\renewcommand{\figurename}{\small{Fig.}~}
\sectionfont{\sffamily\Large}
\subsectionfont{\normalsize}
\subsubsectionfont{\bf}
\setstretch{1.125} 
\setlength{\skip\footins}{0.8cm}
\setlength{\footnotesep}{0.25cm}
\setlength{\jot}{10pt}
\titlespacing*{\section}{0pt}{4pt}{4pt}
\titlespacing*{\subsection}{0pt}{15pt}{1pt}

\fancyfoot{}
\fancyfoot[LO,RE]{\vspace{-7.1pt}\includegraphics[height=9pt]{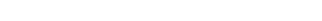}}
\fancyfoot[CO]{\vspace{-7.1pt}\hspace{13.2cm}\includegraphics{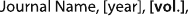}}
\fancyfoot[CE]{\vspace{-7.2pt}\hspace{-14.2cm}\includegraphics{head_foot/RF}}
\fancyfoot[RO]{\footnotesize{\sffamily{1--\pageref{LastPage} ~\textbar  \hspace{2pt}\thepage}}}
\fancyfoot[LE]{\footnotesize{\sffamily{\thepage~\textbar\hspace{3.45cm} 1--\pageref{LastPage}}}}
\fancyhead{}
\renewcommand{\headrulewidth}{0pt} 
\renewcommand{\footrulewidth}{0pt}
\setlength{\arrayrulewidth}{1pt}
\setlength{\columnsep}{6.5mm}
\setlength\bibsep{1pt}

\makeatletter 
\newlength{\figrulesep} 
\setlength{\figrulesep}{0.5\textfloatsep} 

\newcommand{\topfigrule}{\vspace*{-1pt}%
\noindent{\color{cream}\rule[-\figrulesep]{\columnwidth}{1.5pt}} }

\newcommand{\botfigrule}{\vspace*{-2pt}%
\noindent{\color{cream}\rule[\figrulesep]{\columnwidth}{1.5pt}} }

\newcommand{\dblfigrule}{\vspace*{-1pt}%
\noindent{\color{cream}\rule[-\figrulesep]{\textwidth}{1.5pt}} }

\makeatother

\twocolumn[
  \begin{@twocolumnfalse}
{\includegraphics[height=30pt]{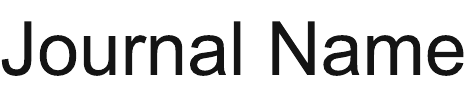}\hfill\raisebox{0pt}[0pt][0pt]{\includegraphics[height=55pt]{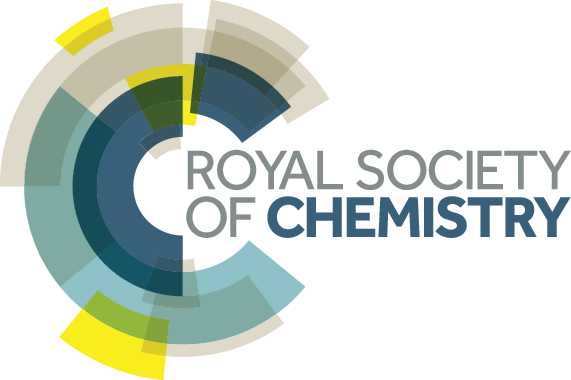}}\\[1ex]
\includegraphics[width=18.5cm]{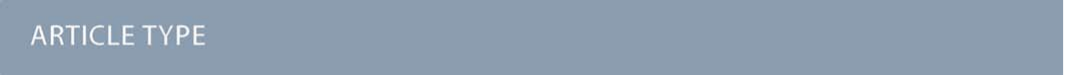}}\par
\vspace{1em}
\sffamily
\begin{tabular}{m{4.5cm} p{13.5cm} }

\includegraphics{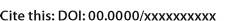} & \noindent\LARGE{\textbf{Tribo-piezoelectric Nanogenerators for Energy Harvesting: a first-principles study}} \\
\vspace{0.3cm} & \vspace{0.3cm} \\

 & \noindent\large{Jemal Yimer Damte$^{\ast}$\textit{$^{a}$} and Jiri Houska\textit{$^{a\ddag}$}} \\

\includegraphics{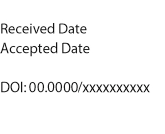} & \noindent\normalsize{Two-dimensional transition metal dichalcogenides (TMDs) are highly promising candidates for various applications due to their unique electrical, optical, mechanical, and chemical properties. Furthermore, heterostructures consisting of TMDs with metals, oxides, and conductive materials have attracted significant research interest due to their exceptional electronic properties. In this study, we utilized density functional theory to investigate those electronic and transport properties, which are relevant for the application of tribo-piezoelectricity in creating novel nanogenerators: an interdisciplinary approach with promising implications. The results of the study demonstrate that the enhancement of charge transfer between layers and the orbital contribution to the Fermi level under applied strain in MoS/IrO, MoS/TiO, MoS/WTe, and MoTe/WS heterostructures is noteworthy. Additionally, non-equilibrium Green's function calculations of electron transport properties provide valuable insights into the behavior of these materials under different conditions. While MoS/IrO and MoS/TiO hetero-bilayers are unsuitable due to their tendency to exhibit large current flow with increasing voltage, others like MoS/WTe and MoTe/WS hetero bilayers show promise due to their ability to prevent voltage drop. The presented innovative concept of utilizing compressive strain of TMD bilayers to generate a tribo-piezoelectric effect for nanogenerators has a potential to contribute to the development of efficient and sustainable energy harvesting devices.
} \\

\end{tabular}

 \end{@twocolumnfalse} \vspace{0.6cm}

  ]

\renewcommand*\rmdefault{bch}\normalfont\upshape
\rmfamily
\section*{}
\vspace{-1cm}


\footnotetext{\textit{$^{a}$~Department of Applied Physics, University of West Bohemia, Univerzitni 2732/8, 301 00 Plzen 3, Pilsen, Czech Republic; E-mail: Jemal Yimer Damte(damtejem@ntis.zcu.cz), Jiri Houska (
jhouska@kfy.zcu.cz)}}



\section{Introduction}
A nanogenerator is a device designed for harvesting small-scale mechanical energy and converting it into electrical energy. The concept behind nanogenerators involves utilizing various physical mechanisms to generate electrical power at the nanoscale\cite{cai2019tribo}.
The field of nanogenerators saw a significant milestone in 2006 with the invention of the first nanogenerator by Zhong Lin Wang et al. This invention marked the beginning of a new era in energy harvesting, particularly at the nanoscale \cite{zi2017nanogenerators, wang2006piezoelectric, fan2012flexible, sripadmanabhan2019nanogenerators}.
Both piezoelectric and triboelectric nanogenerators are prominent techniques in the field of energy harvesting. The triboelectric effect involves the generation of electric charge between two objects due to the contact or slide against each other \cite{liu2019core, liang2015highly}.
On the other hand, the piezoelectric effect involves the generation an electric charge in response to mechanical stress or deformation \cite{won2019flexible, sezer2021comprehensive}.
Researchers have explored a wide range of materials for fabricating piezoelectric nanogenerators, each with its own set of advantages and characteristics such as semiconductors, polycrystalline ceramics, poly vinylidene fluoride, metalloids and a two-dimensional (2D) material \cite{zhou2020all, zaarour2021review, jella2019comprehensive, guan2020hierarchically, chen2019biocompatible}.
The fundamental structure of a piezoelectric nanogenerator typically involves a piezoelectric material sandwiched between metal electrodes.
When subjected to mechanical stress or deformation, piezoelectric materials generate an electric charge \cite{lu2009piezoelectric, wang2014optimization, charoonsuk2019tetragonal}.
 For triboelectric nanogenerator, the triboelectric effect is based on the generation of electric charges due to the relative motion or contact between two different materials \cite{zhu2012triboelectric, wang2012nanoscale, rodrigues2020triboelectric}.

Researchers have explored various strategies to enhance the performance of triboelectric nanogenerators (TENGs) by improving the effective contact area and charge density such as surface functionalization by plasma treatment, liquid-metal electrode application, sponge or foam structure application or silicon template based micro/nano pattern application \cite{yang2016triboelectric, karumuthil2017piezo, shin2015triboelectric, tang2015liquid}.

The integration of piezoelectric and triboelectric effects in a single device, often referred to as a tribo-piezoelectric nanogenerator, offers a synergistic approach to enhance charge density and overall energy harvesting efficiency\cite{zhu2018hybrid, zheng2023design, sriphan2022hybrid}.
 Combining the triboelectric and piezoelectric effects in a single nanogenerator addresses the limitation of low charge density observed in individual piezoelectric or triboelectric nanogenerators. This integration enhances the overall performance and opens up opportunities for various smart applications\cite{nguyen2013effect, sriphan2020multifunctional, pang2021hybrid, lee2022significant}.
In this study, combination of both the triboelectric and the piezoelectric effect in a single material is proposed. This may constitute the next significant advancement in the field of energy harvesting and nanogenerator technology. This innovation addresses challenges related to material selection, simplifying the design process, reducing design complexity, enhancing charge density and optimizing energy harvesting compared to hybrid energy nanogenerators.

Two-dimensional (2D) transition metal dichalcogenides (TMDs like MoS${_2}$, MoSe${_2}$, WS${_2}$, WSe${_2}$, and MoTe${_2}$) have indeed garnered considerable attention in the field of materials science and condensed matter physics. Similar to graphene, TMDs exhibit unique properties due to their two-dimensional nature \cite{wang2014graphene, gmitra2015graphene, rosman2018photocatalytic} .
Semiconducting nature, atomically thin nature, low temperature phenomena and layer dependent properties are some of the key characteristics that make transition metal dichalcogenides (TMDs) crucial in the field of nanotechnology.
Because of its unique properties, TMDs have been explored for various electronic devices, including field-effect transistors, photodetectors, and sensors. \cite{manamela2020electrically, yu2013highly, svatek2021high, asgary2021characterization}.
The use of 2D materials like MoS${_2}$, MoSe${_2}$, and WTe${_2}$ in piezoelectric nanogenerators is a fascinating area of research for harvesting energy due to strain-induced lattice distortion \cite{wu2014piezoelectricity, maity2017two}.
 Futhermore, previous studies investigating the incorporation of two-dimensional (2D) nanomaterials, such as MoS${_2}$, in triboelectric nanogenerators (TENGs) constitute an exciting development in the field of energy harvesting \cite{wu2017enhanced}.
Here, we propose a tribo-piezoelectric device for energy harvesting using different transition metal dichalcogenides. Accordingly, we study the electronic and transport properties of strain and unstrain TMD bilayers using first principle studies. This study can be a guideline for further theoretical and experimental research in this field.

\section{Computational Details} 
In our study, we have considered various hetero-bilayers of MX$_{2}$ materials. The M elements include Mo, W, Ir and Ti, while the X elements include S, Te, and O. The values of strain applied to the hetero-bilayers is reported as $\epsilon$ = $\frac{a - a_{0}}{a_{0}}$, where a and $a_{0}$ represent the strained and unstrained lattice constants, respectively.
 Each model geometry consists of 246 atoms in the unit cell.
All density functional theory (DFT) calculations are performed with the SIESTA package \cite{artacho2008siesta, garcia2010first}, by using the van der Waals density functional(vdW-DF2) exchange functional \cite{klimevs2011van, li2023understanding}.
 After benchmark,  we choose a double polarized basis set, and we fix the kinetic energy cutoff  to 700 eV, and the sampling of the Brillouine zone with a 11x11x1 Monkhorst-Pack mesh.
 A vacuum slab separation is used to eliminate the interaction of neighboring layers.
 All structures are relaxed with conjugate gradient algorithm until the forces on the atoms are less than 0.01 eV/\AA{} and the electronic energy convergence was $10^{-5}$ eV.
The electron transport properties were calculated  self-consistently using the TranSIESTA code which integrates  the non-equilibrium Green's function (NEGF) Method \cite{papior2017improvements}.
The electron transport system includes the left electrode (L), the central scattering region (C), and the right electrode (R).
The semi-infinite left and right electrodes are connected to the the central scattering region.
In the designed tribo-piezoelectric device, gold electrodes are used due to their favorable properties of good ductility and high electrical conductivity.
The electric current flows across the device under the applied voltage.
 The Landauer–Büttiker formula used to calculated the electric current \cite{buttiker1985generalized} which can be found from the integration of the transmission curve as follows:
 \begin{equation}
\frac{2e}{h}\int_{-\infty}^{\infty}
{T(E, V_{b})[f_{L}(E-\mu_{L})-f_{R}(E-\mu_{R})]dE} = I(V_{b})
\end{equation}

where $f_{L(R)}$, e, $I(V_{b})$, $\mu_{L(R)}$ and h represent the Fermi Dirac distribution function of the left and right electrode, the electron charge, the electric current under applied bias voltages, the chemical potential of the left and right electrode and the Planck's constant, respectively.
The chemical potential of the two electrodes ($\mu_{L(R)} = E_{F}\pm \frac{eV_{b}}{2})$ is shifted up and down from the Fermi energy $E_{F}$, and T(E,$V_{b})$ is the transmission coefficient:
\begin{equation}
T(E,V_{b}) = Tr[\Gamma_{L} (E,V_{b})G(E,V_{b})\Gamma_{R}(E,V_{b})G^\star(E,V_{b})]
\end{equation}

The retarded and advanced Green's function are represented by G(E,$V_{b}$) and $G^\star(E,V_{b})$, respectively. $\Gamma_{L}$ is the contact broadening function with respect to the left electrode and
$\Gamma_{R}$ is the contact broadening function with respect to the right electrode \cite{shin2011interflake, yeh2022combined}.
Finally, post-processing tool TBTrans included in the TranSIESTA package was used to calculate the electron transmission function and the electric current.
The current-voltage characteristic of each system  was obtained at different bias conditions.
 Various electronic analyses, such as Bader charges, projected density of states and charge density calculations, have been performed to understand the electronic properties of the materials. Additionally, electronic transport calculations has been carried out to identify the best tribo-piezoelectric materials.
 \begin{figure}[h]
\centering
  \includegraphics[height=8cm]{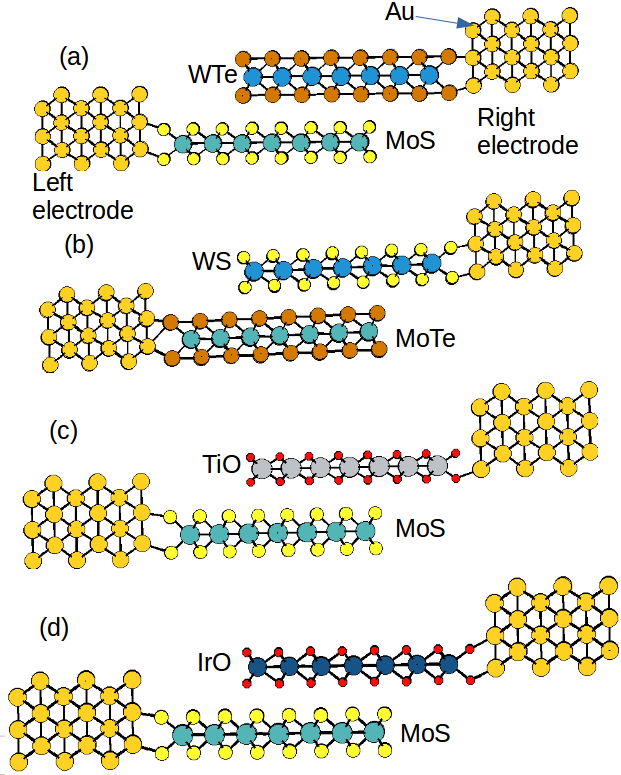}
  \caption{Hetero-bilayers of (a)MoS-WTe/Au system, (b)MoTe-WS/Au system, (c)MoS-TiO/Au system and (d)MoS-IrO/Au system (layer 1 is the bottom layer and layer 2 is the top layer in each heterostructure).}
  \label{fgr:1}
\end{figure}
                                                                                                                                                                                                                                                                                                                                                                                                                
\section{Results and discussion}
\subsection{Bader charge analysis}
The grid-based Bader analysis algorithm is a method used to divide the charge density grid into Bader volumes. It involves identifying the zero flux surface, which separates one Bader volume from another. Each Bader volume contains the Bader charge associated with a specific atom.
 To calculate the charges on layers, a post-processing analysis is performed using the on-grid Bader analysis method developed by the Henkelman group and then the charge density gradient in the direction
 $\hat{r}$, is calculated as follows \cite{henkelman2006fast, tang2009grid, sanville2007improved}:

\begin{equation}
\Delta\rho.\hat{r} = \frac{\Delta\rho}{|\Delta \vec{r}|}
\end{equation}
                                                                                                                                                                                                         where $\Delta\rho$ is the change in charge density and $\Delta \vec{r}$ is the change of distance.
The total charge is obtained by summing over all the atoms in the system.
To determine the charges on layer 1 (the bottom layer), layer 2 (the top layer), and the electrodes, unstrained structure is used for comparison. The charges on layer 1, layer 2, and the electrodes are obtained by subtracting the charges of the heterostructures from the charges of the unstarined structure.
 Positive values indicate a loss of electrons, while negative values imply a gain of electrons.
 The Bader charge analysis results for different heterostructures are presented in Table 1. The Bader charge analysis shows that layer 1 accepts a certain number of electrons, while layer 2 loses a corresponding number of electrons. The specific amounts of charge transfer depend on the particular heterostructure. The results shows that in MoS/TiO and MoS/WTe heterostructures, layer 1 gains electrons, while layer 2 loses electrons (Table 1). 
 In the case of MoS/IrO and MoTe/WS heterostructures, layer 1 loses electrons, while layer 2 gains electrons.
The Bader charge analysis provides valuable insights into the charge transfer between layers in various heterostructures. The observation of increased charge transfer in MoS/IrO, MoS/TiO, MoS/WTe, and MoTe/WS heterostructures under higher applied strain  ${-4}$\%  compared to a lower strain $ {-2}$\%  is noteworthy.
From this table, it is observed that under ${-4}$\% applied strain, MoS/IrO, MoS/TiO, MoS/WTe, and MoTe/WS heterostructures exhibit a slightly higher amount of charge transfer compared to $ {-2}$\% applied strain. The increase in charge transfer under higher applied strain indicates a strain-induced redistribution of charges between the layers. This could be attributed to changes in the interlayer distances, atomic positions, or electronic configurations caused by strain. The fact that MoS/IrO, MoS/TiO, MoS/WTe, and MoTe/WS heterostructures exhibit slightly higher charge transfer under increased strain suggests that these specific combinations of materials are particularly responsive to mechanical deformation.
 Understanding the nature of charge transfer and the electronic interactions between layers provides insights into how these heterostructures respond to strain. This information is crucial for applications such as nanogenerators, where mechanical stress is harnessed for energy harvesting.

\begin{table}
\begin{center}
\caption{Calculated Bader charge values of (MoS-WTe, MoTe-WS and MoS-TiO and MoS-IrO)/Au system at different strains.}

\begin{tabular}{c c c c c c}
 \hline
 MoS/WTe & Layers & ${-2}$\% & ${-4}$\% \\ [0.5ex]
 \hline
 Layer 1 & MoS &  -0.09 &  -0.10 \\

 Layer 2 & WTe &  0.06 &  0.12 \\

 Electrode & Right & 0.01 &  -0.04 \\

 Electrode  & Left & 0.02  &  0.03 \\ [1ex]
 \hline
 MoTe/WS & Layers & ${-2}$\% & ${-4}$\% \\ [0.5ex]
 \hline
 Layer 1 & MoTe &  0.04 &  0.05 \\

 Layer 2 & WS &  -0.06 &  -0.08 \\

 Electrode & Right & 0.01 & 0.01 \\

 Electrode  & Left & 0.02 &  0.03 \\ [1ex]
 \hline
 MoS/TiO & Layers & ${-2}$\% & ${-4}$\% \\ [0.5ex]
 \hline
 Layer 1 & MoS & -0.06 &  -0.10 \\

 Layer 2 & TiO &  0.05 &  0.06 \\

 Electrode & Right & -0.01 & 0.01 \\

 Electrode  & Left & 0.01 &  0.02 \\ [1ex]
 \hline
 MoS/IrO &Layers & ${-2}$\% & ${-4}$\% \\ [0.5ex]
 \hline
 Layer 1 & MoS & 0.06 &  0.13 \\

 Layer 2 & IrO &  -0.05 &  -0.10 \\

 Electrode & Right & -0.01 &   -0.02 \\

 Electrode  & Left & -0.02 &  -0.01  \\ [1ex]
 \hline
\end{tabular}
\end{center}
 \end{table}
                                                                                                                                                                                                                                                                                                                                                                                                                                                                                                                                                                                                                                                                                                                                                                                                                                                                                                                                                                                                                                                                                                                                                                                                                                                                                                                                                                                                                                                                                                                                                                                                                                                                                                                                                                                                                                                                                                                                                                                                                                                                                                                                                                                                                                                                                                                                                                                                                                                                                                       
\subsection{Density of states and charge density}

Density of states measures the number of states at a particular energy levels per unit volume for electrons in the conduction band and holes in the valance band. 
Therefore, the projected density of states (PDOS) of all heterostructures has been investigated to understand the interaction of layers under applied strain.
The observation of slight shifts in the d-orbitals of Mo toward the Fermi level with increasing strain, as depicted in the PDOS plots (Figure 2 and 3), constitutes important insight into the electronic structure of the heterostructures. The slight shift of Mo d-orbitals toward the Fermi level with increasing strain suggests that the applied mechanical deformation affects the electronic orbitals of Mo. This could be due to changes in the interatomic distances or electronic configurations caused by strain. In addition, the observation of minor changes in the band gap under applied strain is also noteworthy. The band gap is a key parameter that determines the material's electrical conductivity. Even minor alterations in the band gap can have significant implications for the material's behavior. The observed electronic structure changes under strain suggest that these heterostructures could find applications in devices that leverage strain sensitivity, such as strain sensors, actuators, or devices based on the piezoelectric effect.

In summary, the slight shifts in Mo d-orbitals and minor changes in the band gap under applied strain highlight the intricate interplay between mechanical deformation and electronic properties in the heterostructures. These findings open avenues for exploring the strain-induced tunability of electronic characteristics for various applications.
\begin{figure}[h]
\centering
  \includegraphics[height=6cm]{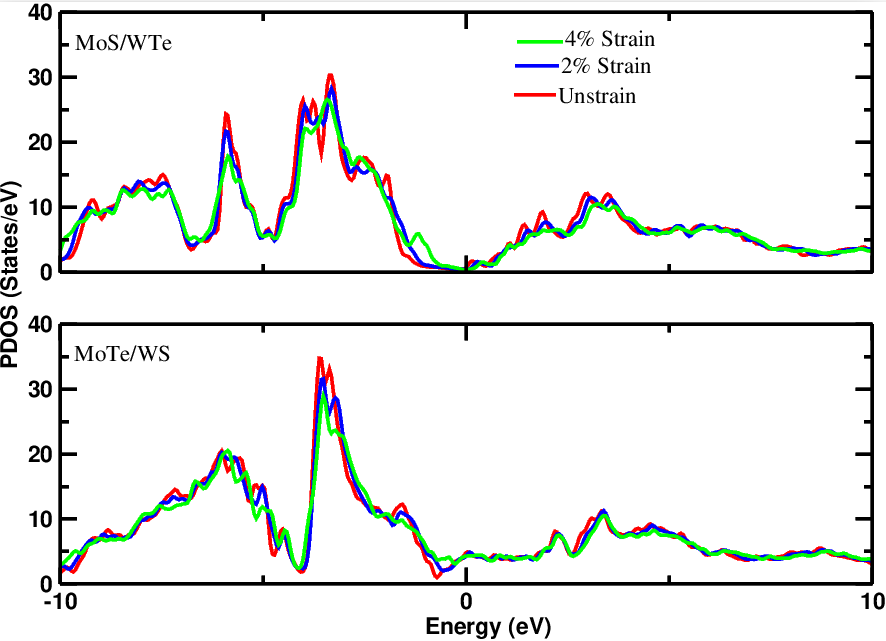}
  \caption{Projected density of states d-orbitals of Mo in MoS-WTe/Au system and MoTe-WS/Au system at different strains (${-2}$\% and ${-4}$\%).The fermi level is set to zero.}
  \label{fgr:2}
\end{figure}

\begin{figure}[h]
\centering
  \includegraphics[height=6cm]{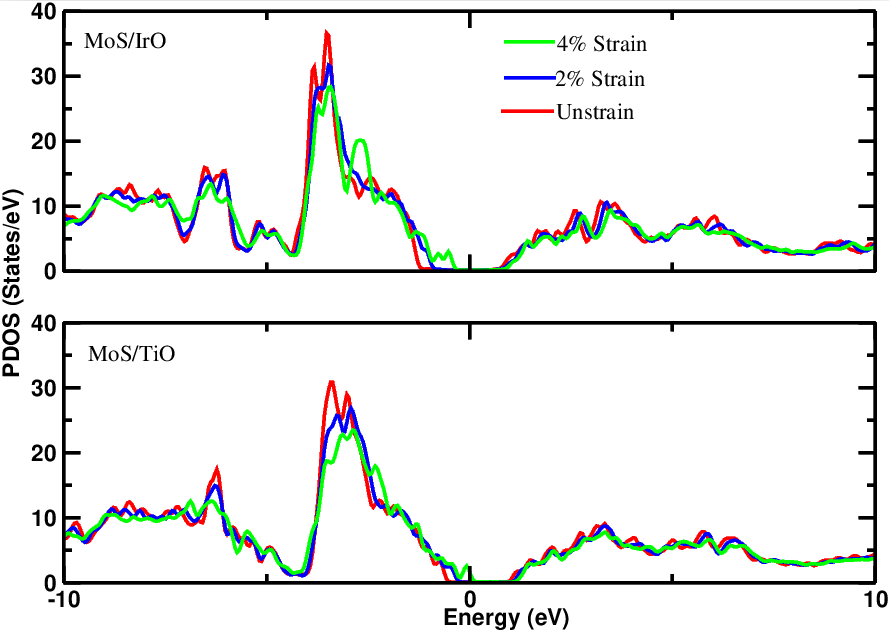}
  \caption{Projected density of states d-orbitals of Mo in MoS-IrO/Au system and MoS-TiO/Au system at different strains (${-2}$\% and ${-4}$\%). The Fermi level is set to zero.}
  \label{fgr:3}
\end{figure}

The electronic charge density analysis provides valuable insights into the interaction of layers in heterostructures. The charge density plots shown in Figure 4 depict the distribution of electrons around the atoms in MoTe/WS hetero bilayers.
 In these plots, the color map serves to differentiate between various
levels of electron localization and delocalization. Specifically, the
pink color signifies regions with higher charge density, indicating a
greater concentration of electrons. On the other hand, the blue color
indicates electron delocalization, suggesting that electrons are more
spread out in these regions. The cyan color indicates minimal delocalization of electrons, suggesting a weaker interaction or bonding between them, while the red and yellow colors represent regions where there is no significant charge density available.
 \begin{figure}[h]
\centering
  \includegraphics[height=2.5cm]{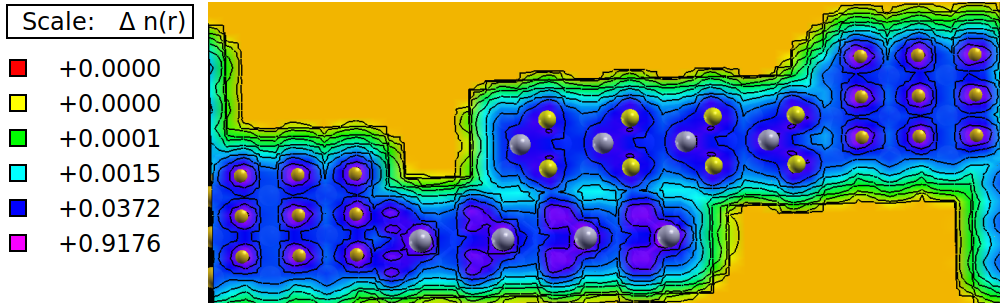}
  \caption{Charge density plot of MoTe/WS hetero bilayer, isosurface value is 0.05 electron/${bohr}^3$.}
  \label{fgr:4}
\end{figure}

Analyzing the charge density plots, it can be observed that between the layers the electrons are slightly localized, as indicated by the blue color scheme between the layers. This suggests a certain degree of electron confinement or bonding between the layers.
Similarly, within the atomic bonds, there is a slight electron localization.

\subsection{ Current-Voltage Characteristics}

A triboelectric nanogenerator involves the contact electrification between two dielectric layers, which are separated by a gap and connected to electrodes. 
This process generates electrostatic charges with opposite signs, resulting from the contact. If we assume that the planar size of the dielectric film is much larger than their separation distance, we can calculate the electric field strength between them as follows: \cite{wang2020first, niu2013theoretical}.

\begin{equation}
E_{z} = -\frac{\sigma(z,t)}{\varepsilon_{1}},
\end{equation}
\begin{equation}
E_{z} = -\frac{\sigma(z,t)}{\varepsilon_{2}} ,
\end{equation}
\begin{equation}
E_{z} = -\frac{(\sigma(z,t)-\sigma_{T})}{\varepsilon_{0}}.
\end{equation}

The potential drop between two electrodes is calculated as follows:
\begin{equation}
\phi = -\sigma(z,t)[\frac{d_{1}}{\varepsilon_{1}}+ \frac{d_{2}}{\varepsilon_{2}}]- H(t)\frac{[\sigma(z,t)-\sigma_{T}]}{\varepsilon_{0}}
\end{equation}
where $\sigma_{T} (t)$, $\sigma (z,t)$, $\varepsilon$, H and d is surface charge density, the density of free electrons on surfaces of the electrodes, permitivity, a function of time and thickness, respectively.
The current-voltage (I-V) characteristics were studied using the norm equilibrium Green's function (NEGF) technique at various bias voltage ranges (0.0 - 1.0 V). At each bias voltage, the Landauer-Büttiker formula was utilized with a 0.1 V interval.
 For the transport setup, Au electrodes were employed on both sides, serving as the left and right electrodes. These electrodes were assumed to be semi-infinite. The scattering region, where the heterostructures were located, was positioned between the left and right electrodes and connected to them in the transport calculations. 
 Figure 5 illustrates the I-V characteristics of the MoS/WTe, MoTe/WS, MoS/TiO and MoS/IrO heterostructures. The current response exhibits minor changes and remains within the same order of magnitude in the MoS/WTe and MoTe/WS heterostructures. However, the current response increases as the voltage is raised under applied strain. This behavior indicates rectifying I-V characteristics and suggests an improvement in transmission performance, which is consistent with previous research \cite{zhou2019influence, sun2017first}.
 
  \begin{figure}[h]
\centering
  \includegraphics[height=7cm]{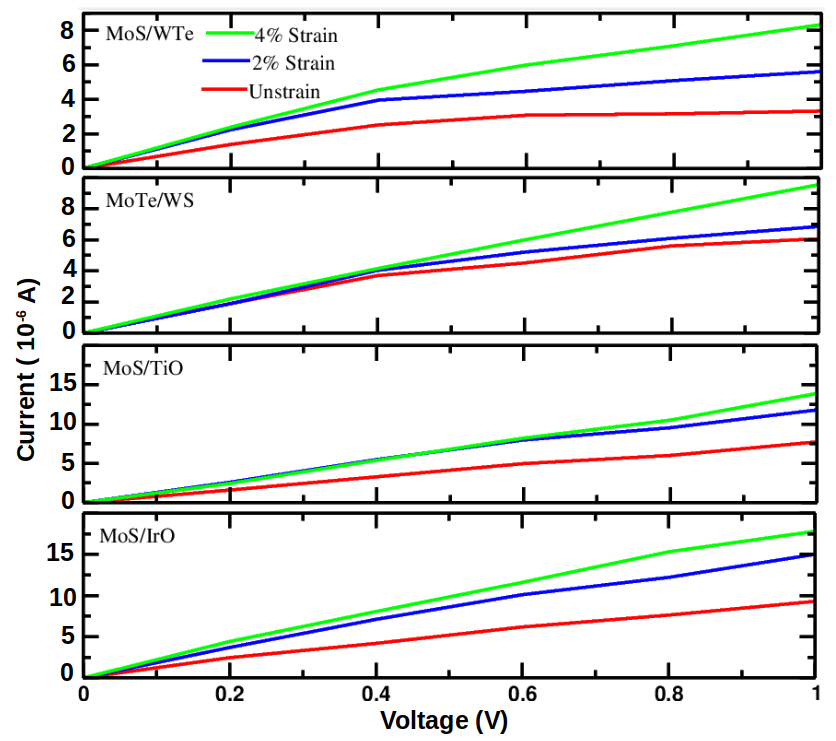}
  \caption{Current-voltage characteristic of various TMD bilayers at different strains (${-2}$\% and ${-4}$\%).}
  \label{fgr:5}
\end{figure}

 The I-V curves demonstrate that the current value increases with the increment of the bias voltage, confirming enhanced electron transport transmission across the scattering center.
 Similarly, Figure 5 displays the I-V characteristics of the MoS/TiO and MoS/IrO heterostructures, which involve the integration of two-dimensional metal dichalcogenides with other transition metal oxides.
For example, a current which is (1) larger than $10^{-6}$ A and at the same time (2) at least $\approx$2x larger compared to the unstrained bilayer is predicted in the MoS/TiO heterostructure at the $ {-2}$\% strain and a 1.0 V bias voltage or at the $ {-4}$\% strain and a 0.8-1.0 V bias voltage, and in the MoS/IrO heterostructure even at a 0.6-1.0 V bias voltage regardless the strain. On the one hand, the MoS/WTe and MoTe/WS heterostructures exhibit lower current response, and the latter one also weaker enhancement of the current response by the strain applied.
The substantial enhancement in current response under strain and bias voltage variations suggests that MoS/TiO and MoS/IrO heterostructures could find applications in devices requiring sensitivity to mechanical strain or specific bias conditions. In summary, MoS/IrO and MoS/TiO hetero-bilayers are unsuitable due to their tendency to exhibit large current flow with increasing voltage, while others like MoS/WTe and MoTe/WS hetero bilayers show promise due to their ability to prevent voltage drop.
The observed variations in current response provide opportunities for optimizing device performance by tailoring strain, bias voltage, or other environmental conditions. Understanding the interplay between these factors is crucial for designing devices with desired electrical characteristics. 
Furthermore, it is worth noting that different heterostructures exhibit distinct I-V characteristics, which aligns with findings from previous studies \cite{zhou2019influence}.
The significant enhancement in current response under strain and bias voltage variations in the selected heterostructures opens avenues for applications in strain-sensitive and voltage-controlled devices, respectively. These findings contribute to the growing field of heterostructure-based electronic devices with tunable and optimized performance.

\section{Conclusions}
We have carried out density functional theory (DFT) studies to investigate the Bader charge analysis, density of states and transport properties of different hetero-bilayers. Bader charge analysis was conducted to understand the charge transfer between layers in hetero-bilayers under applied strain. The results indicate that the charge transfer between layers increases under applied strain. The interactions of layers in heterostructures under applied strain have been examined. A more significant contribution to the Fermi level has been revealed in strained heterostructures compared to unstrained ones. We have utilized TranSIESTA, a component of the SIESTA package, to calculate electron transport properties. An increase in current response with the applied voltage, with a more significant response under applied strain compared to the unstrained heterostructure has been observed. While MoS/IrO and MoS/TiO hetero-bilayers are unsuitable due to their tendency to exhibit large current flow with increasing voltage, others like MoS/WTe and MoTe/WS hetero bilayers show promise due to their ability to prevent voltage drop. Based on the tribo-piezoelectricity observed in TMD bilayers, the compression strain is suggested as a means to enhance the performance of nanogenerators. We believe that that our findings provide guidelines for further theoretical and experimental research in this field.

\section*{Author Contributions}
JYD: Conceptualization, Data curation, Investigation, Methodology, Visualization, Writing - orginal draft
JH: Writing - review and editing

\section*{Conflicts of interest}
There are no conflicts of interest to declare.

\section*{Acknowledgements}

This work was supported by the project Quantum materials for applications in
sustainable technologies (QM4ST), funded as project No. CZ.$02.01.01\slash 00\slash 22{\_}008\slash 0004572$ by Programme Johannes Amos Commenius, call Excellent Research.



\balance


\bibliography{rsc} 
\bibliographystyle{rsc} 

\section*{Graphical abstract}

{\centering
\includegraphics[height=5cm]{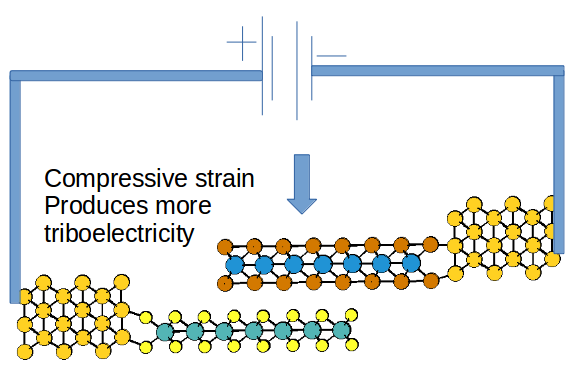}
\par
}

\end{document}